\documentclass[letter,final]{IEEEtran}
\usepackage[utf8]{inputenc}
\usepackage{graphicx}
\usepackage{amssymb}
\usepackage[cmex10]{amsmath}
\usepackage{color}
\usepackage{theorem}
\usepackage{pifont}
\usepackage{hyperref}
\usepackage{url}
\usepackage{breakurl}
\usepackage{nicefrac}
\usepackage{booktabs}
\usepackage{bbm}

\usepackage[english]{babel}
\usepackage[english = american]{csquotes}
\usepackage{mathtools}
\usepackage{mathdots}
\MakeOuterQuote{"} 
\usepackage[ruled,vlined,titlenumbered]{algorithm2e}

\definecolor{light_gray}{rgb}{0.6,0.6,0.6}
\definecolor{awgray}{rgb}{0.7,0.7,0.7}
\definecolor{awgray_dark}{rgb} {0.4,0.4,0.4}

\usepackage{pgf}
\usepackage{tikz}
\usetikzlibrary{matrix,chains,positioning,arrows,calc,decorations,plotmarks,patterns, fit,backgrounds}
\usepackage{pgfplots}
\usepgfplotslibrary{external}
\usetikzlibrary{external}                       
\tikzsetexternalprefix{tikzpic/}                   

\pgfplotscreateplotcyclelist{mylist}{red,blue,black,yellow,brown}

\tikzset{
    >=stealth',
    mycircle/.style={circle, draw=black, thick, text width=.1em, minimum height=.8em, text centered, font=\scriptsize}, 
    mycircle_gray/.style={circle, draw=gray, thick, text width=.1em, minimum height=.8em, text centered, font=\tiny},    
    mycircle_small/.style={circle,draw=awgray_dark,fill = awgray_dark, inner sep=0,minimum size=.6em},       
    mycircle_small_black/.style={circle,draw=black,fill = black, inner sep=0,minimum size=.6em},   
    mybox/.style={rectangle,rounded corners,draw=black, thick,text width=1em,minimum height=4em,minimum width=4em,text centered},     
    mybox_small/.style={rectangle,rounded corners,draw=black, thick,text width=1em,minimum height=2em,minimum width=2em,text centered},               
    mybox_vec/.style={rectangle,rounded corners,draw=black, thick,text width=1em,minimum height=0.7em, minimum width=4em,text centered},  
    mybox_vec_short/.style={rectangle,rounded corners,draw=black, thick,text width=1em,minimum height=0.7em, minimum width=2em,text centered},                  
    pil/.style={->, thick, shorten <=2pt, shorten >=2pt,},
}

\usepackage[ruled,vlined,titlenumbered]{algorithm2e}
\makeatletter
\newcommand{\removelatexerror}{\let\@latex@error\@gobble}
\makeatother

\IEEEoverridecommandlockouts

\newtheorem{theorem}{Theorem}
\newtheorem{definition}{Definition}

\newtheorem{corollary}{Corollary}

\newtheorem{example}{Example}

 \newcommand{\qed}{\hfill \mbox{\raggedright \rule{.07in}{.1in}}}

\newcommand{\Fq}{\ensuremath{\mathbb F_{q}}}

\newcommand{\FTwo}{\ensuremath{\mathbb F}_2}
\newcommand{\myset}[1]{\mathcal{#1}}
\newcommand{\intervallincl}[2]{\ensuremath{[#1,#2]}}

\SetAlgoCaptionSeparator{.} 

\newcommand{\printalgoIEEE}[1]
{\begin{center}
\vspace{1ex}
\scalebox{0.95}{
\begin{tabular}{p{.5\textwidth}}
\removelatexerror
\begin{algorithm}[H]
 #1
\end{algorithm}
\end{tabular}
}
\vspace{1ex}
\end{center}
}

\DeclareMathOperator{\wt}{wt}

\newcommand{\wtH}[1]{\wt_{\fontmetric{H}}(#1)}

\renewcommand{\vec}[1]{\ensuremath{\textup{\textbf{#1}}}}

\renewcommand{\c}{\vec{c}}

\renewcommand{\r}{\vec{r}}

\newcommand{\mycode}[1]{\ensuremath{\mathcal{#1}}}

\newcommand{\fontmetric}[1]{\mathsf{#1}}

\newcommand{\codeID}[1]{\ensuremath{(#1)^{\fontmetric{L}}_q}}

\newcommand{\dID}{\ensuremath{d_{\fontmetric{L}}}}

\newcommand{\tauID}{\ensuremath{\tau_{\fontmetric{L}}}}

\newcommand{\ellID}{\ensuremath{\ell_{\fontmetric{L}}}}

\newcommand{\BallID}[2]{\mathcal{B}^{(#1)}_{\fontmetric{L}}(#2)}
\newcommand{\BallIn}[2]{\mathcal{B}^{(#1)}_{\fontmetric{I}}(#2)}
\newcommand{\BallD}[2]{\mathcal{B}^{(#1)}_{\fontmetric{D}}(#2)}
\newcommand{\BallI}[2]{\mathcal{B}^{(#1)}_{\fontmetric{I}}(#2)}

\begin{document}

\title{List Decoding of Insertions and Deletions}

\author{Antonia Wachter-Zeh, \IEEEmembership{Member IEEE}
		\thanks{
				This work was supported by the Technical University of Munich---Institute for Advanced Study, funded by the German Excellence Initiative and European Union Seventh Framework Programme under Grant Agreement No.~291763.
				
				Parts of these results were presented at the \emph{IEEE International Symposium on Information Theory (ISIT) 2017, Aachen, Germany~\cite{Wachterzeh-ISIT2017}.}
			
			A. Wachter-Zeh is with the Institute for Communications Engineering, Technical University of Munich (TUM), Munich, Germany, Email: {antonia.wachter-zeh@tum.de}.
		}}
\maketitle

\begin{abstract}
List decoding of insertions and deletions in the Levenshtein metric is considered.
The Levenshtein distance between two sequences is the minimum number of insertions and deletions needed to turn one of the sequences into the other.
In this paper, a Johnson-like upper bound on the maximum list size when list decoding in the Levenshtein metric is derived. This bound depends only on the length and minimum Levenshtein distance of the code, the length of the received word, and the alphabet size. It shows that polynomial-time list decoding beyond half the Levenshtein distance is possible for many parameters. 
Further, we also prove a lower bound on list decoding of deletions with with the well-known binary Varshamov-Tenengolts (VT) codes which shows that the maximum list size grows exponentially with the number of deletions.
Finally, an efficient list decoding algorithm for two insertions/deletions with VT codes is given.
This decoder can be modified to a polynomial-time list decoder of any constant number of insertions/deletions.
\end{abstract}


\section{Introduction}\label{sec:introduction}
Lately, codes for correcting insertions and deletions attract a lot of attention due to their possible application to correcting errors in DNA storage and DNA sequencing, cf. \cite{JainFarnoudSchwartzBruck-DuplicationDNA}.
Furthermore, insertions and deletions can be seen as synchronization errors in communication systems since the information about the position is lost when receiving a certain symbol.

The algebraic concepts by Varshamov, Tenengolts, and Levenshtein for correcting insertions and deletions go back to the 1960s. Varshamov and Tenengolts designed a class of codes, nowadays called \emph{VT codes}, which was originally designed to correct asymmetric errors on the Z-channel~\cite{tenengolts1984nonbinary,VarshTene-SingleDeletion1965}. Levenshtein proved that these codes can also be used to correct a single insertion or deletion~\cite{Levenshtein-binarycodesCorrectingDeletions}. Further, he constructed a class of codes that can correct two adjacent insertions or deletions~\cite{Levenshtein-TwoAdjacentDeletions}, and designed $q$-ary single insertion-/deletion-correcting codes~\cite{tenengolts1984nonbinary}.
Brakensiek, Guruswami and Zbarsky recently presented multiple insertion-/deletion-correcting codes in~\cite{BrakensiekGuruswamiZbarksy-DeletionCorrection} with low redundancy.
The papers \cite{ChengSwartFerreiraAbdelGhaffar-ThreeOrMoreBurstDelIns,BurstDeletions-ISIT,SchoenyWachterzehGabrysYaakobi-BurstDeletions-journal} deal with bursts of insertions/deletions.
Special types of insertions include duplications \cite{LenzWachterzehYaakob-WCC2017} and sticky insertions \cite{Mitzenmacher-StickyInsertions}.
Asymptotic bounds on the size of insertion-/deletion-correcting codes were derived by Levenshtein~\cite{Levenshtein-binarycodesCorrectingDeletions} and non-asymptotic bounds 
were recently presented by Kulkarni and Kiyavash~\cite{KulkarniKiyavash-UpperBoundsForDeletion}.

This paper investigates \emph{list decoding} of insertions and deletions. 
A \emph{list decoder} returns the list of all codewords in radius at most $\tau$ around a given received word. 
The idea of list decoding was introduced by Elias \cite{elias_list_1957} and Wozencraft \cite{wozencraft_list_1958}.
In this paper, we investigate list decoding in the Levenshtein metric.
The Levenshtein distance between two words is the minimum number of insertions and deletions needed to turn one of the words into the other. 
A $(\tau,\ell)_\mathsf{L}$-\emph{list decoder} is therefore a decoder which returns all codewords of a given code in Levenshtein distance at most~$\tau$ around any given received word. The size of this list is at most $\ell$.

A fundamental question when considering list decoding in a certain metric is whether the complexity of a list decoding algorithm is "feasible". This leads to investigating the possibility of \emph{polynomial-time} (w.r.t. the code-length) list decoding.
The list size $\ell$ gives a lower bound on the complexity of such a decoder, since writing down the whole list of codewords is part of any list decoding algorithm.
On the one hand, a lower bound on the maximum list size that grows exponentially in the length of the code rules out the possibility of polynomial-time list decoding.
This is the case for example for certain families of rank-metric codes, cf. \cite{Wachterzeh_BoundsListDecodingRankMetric_IEEE-IT_2013,RavivWachterzeh_GabidulinBounds_journal}.
On the other hand, a polynomial upper bound shows that polynomial-time list decoding is possible (but explicit algorithms still have to be found).
In the Hamming metric, the \emph{Johnson upper bound} \cite{Johnson1962New,Johnson1963Improved,Bassalygo1965New,Guruswami_ListDecodingofError-CorrectingCodes_1999} shows for \emph{any} code of length $n$ and minimum Hamming distance $D$ that the size of the decoding list is polynomial in $n$ when $\tau$ is less than the so-called Johnson radius $n- \sqrt{n(n-D)}$.

In this paper, we investigate list decoding of insertions and deletions in the Levenshtein metric. First, we derive a Johnson-like upper bound on the list size. 
Second, a lower bound on list decoding VT codes and a list decoding algorithm of a constant number of insertions/deletions with VT codes are given.

To our knowledge, this is the first paper which considers list decoding of insertions/deletions in the Levenshtein metric. The recent work by Mazooji~\cite{Mazooj-UniqueDecodingInsertions_ISIT,Mazooj-UniqueDecodingInsertionsBeyond_2016} considers \emph{unique} decoding beyond half the minimum Levenshtein distance, but this is a different setting since such a decoder always fails when the list size is larger than one. 
The paper \cite{GuruswamiWang-HighNoiseHighRateDeletions_2017} by Guruswami and Wang shows that there exists a binary code of length $n$ which can list-decode $n\left(\frac{1}{2}-\epsilon\right)$ deletions, where $0 < \epsilon <\frac{1}{2}$. However, they consider \emph{deletions only}, and their code rate is very small, in the order of $\epsilon^3$. In this region, our upper bound shows that from a combinatorial point of view, polynomial-time list decoding is actually possible up to a larger radius.
Another similar-looking problem is the \emph{sequence reconstruction problem} of insertions or deletions, see \cite{L01B,SalaGabrysSchoenyDolecek-ReconstructionsInsertions,GY16}, where the maximum intersection of two insertion/deletion balls of radius $\tau$ around any two words is calculated. The problem in this paper can on the contrary be seen as calculating the maximum number of \emph{code}words in a ball around any word.

The rest of this paper is organized as follows. Section~\ref{sec:prelim} provides notations and definitions, and defines VT codes. In Section~\ref{sec:johnson}, we derive the main result, i.e., the Johnson-like upper bound on the maximum list size and also give an alphabet-free version of this bound. This bound is then illustrated and discussed in Section~\ref{sec:discussion} and simplified for the cases of deletions only and insertions only. Section~\ref{sec:lower-bound} provides the lower bound on list decoding deletions only with VT codes and finally, in Section~\ref{sec:list-dec-vt} we give a list decoder for two insertions/deletions with VT codes and explain how it can be extended to any constant number of insertions/deletions.

\section{Preliminaries}\label{sec:prelim}
\subsection{Notations and Definitions}
Let $\Fq$ be a finite field of order $q$, where $q$ is a power of a prime, and let $\Fq^n$ denote the set of all vectors (sequences) of length $n$ over $\Fq$.
A \emph{subsequence} of a vector $\vec{x} = (x_1,x_2,\dots,x_n)$ is formed by taking a subset of the symbols of $\vec{x}$ and aligning them without changing their order. Hence, for $ N<n $, any vector $\vec{y} = (x_{i_1},x_{i_2}, \dots, x_{i_{N}} )$ is a subsequence of $\vec{x}$ if $1 \leq i_1 < i_2 < \dots < i_N \leq n$, and in this case we say that $n-N$ \emph{deletions} occurred in the sequence~$\vec{x}$ and $\vec{y}$ is the result.

Vice versa, $N-n$ \emph{insertions} occurred in the sequence~$\vec{x}$ if $N >n$ and $\vec{x}$ is a subsequence of $\vec{y} = (y_1,y_2,\dots,y_N)$. 
Here, $\vec{y}$ is called a \emph{supersequence} of~$\vec{x}$.




The Levenshtein distance $\dID(\vec{a},\vec{b})$ between two words $\vec{a}$ and~$\vec{b}$ (not necessarily of the same length) is the minimum number of insertions and deletions which is needed to transform $\vec{a}$ into $\vec{b}$. In this sense, an additive error of Hamming weight one (also called substitution error) consists of one insertion and one deletion at the same position. Therefore, if the Hamming distance between $\vec{a}$ and $\vec{b}$ is $d_{\mathsf{H}}(\vec{a}, \vec{b})= D$, then $\dID(\vec{a}, \vec{b}) \leq 2D$.
Also notice that for two vectors $\vec{a}$ of length $n$ and $\vec{b}$ of length $N$, $\dID(\vec{a}, \vec{b})$ can become larger than $n$ and at most $n+N$.

By $\BallID{\tau}{\vec{r}}$ we denote the set of all vectors in Levenshtein distance at most $\tau$ around a given word $\vec{r}$; by $\BallD{\delta}{\vec{r}}$ the set of all vectors which can be obtained from $\vec{r}$ by at most~$\delta$ deletions; and by $\BallI{\epsilon}{\vec{r}}$ the set of words which can be obtained from $\vec{r}$ by at most $\epsilon$ insertions.
The size of $\BallID{\tau}{\vec{r}}$ and $\BallD{\delta}{\vec{r}}$ depends on its \emph{center}, whereas the size of $\BallI{\epsilon}{\vec{r}}$ only depends on the radius $\epsilon$ and \emph{the length $N$} of $\mathbf{r}$ and therefore, by slight abuse of notation, we also write $|\BallI{\epsilon}{N}|$.

By the triple $\codeID{n,M,d}$ we denote a code of Levenshtein distance 
$d$, of length $n$, and cardinality $M$.
An $\codeID{n,M,d}$ code can always decode uniquely if the number of insertions $\epsilon$ and the number of deletions $\delta$ satisfy
\begin{equation}\label{eq:unique-dec}
\tau \triangleq\epsilon + \delta < \frac{d}{2}.
\end{equation}
In this paper, we analyze the capability of $\codeID{n,M,d}$ codes to perform polynomial-time list decoding beyond $d/2$.
Given a received word $\vec{r} \in \Fq^N$, a $(\tau,\ell)_\mathsf{L}$-\emph{list decoder} for an $\codeID{n,M,d}$ code returns the set of all codewords in Levenshtein distance at most $\tau$ from $\vec{r}$ and the size of this list is at most~$\ell$.


Throughout this paper, the \emph{maximum list size} is denoted by $$\ell\triangleq \max_{\vec{r} \in \Fq^{N}}\big\{\big|\mycode{C} \cap \BallID{\tau}{\vec{r}}\big|\big\}.$$

Finally, for any positive integers $a,b$, we denote by $[a,b]$ the set of integers $\{i:a \leq i \leq b, i \in \mathbb{Z}\}$.
\subsection{Varshamov--Tenengolts Codes}
The binary \emph{Varshamov--Tenengolts} (VT) codes \cite{VarshTene-SingleDeletion1965} are single-deletion correcting codes (cf. Sloane's survey in~\cite{Sloane01onsingle-deletion-correcting}), defined as follows.
\begin{definition}[VT Codes]\label{def:vt-code}
	For $a \in [0,n]$, the binary Varshamov--Tenengolts (VT) code $\mathcal{VT}_{a}(n)$ is the following set of binary vectors:
	\begin{equation*}
	\mathcal{VT}_{a}(n) \triangleq \Big\{\vec{c}=(c_1,\ldots,c_n) \, : \, \sum_{i=1}^{n}i \cdot c_i \equiv a~\bmod (n+1)\Big\}.
	\end{equation*}
\end{definition}
Levenshtein proved in~\cite{Levenshtein-binarycodesCorrectingDeletions} that VT codes can correct either a single deletion or a single insertion. The largest VT codes are obtained for $a=0$, and these codes are conjectured to be optimal in the sense that they have the largest cardinality among all single-deletion correcting codes~\cite{Sloane01onsingle-deletion-correcting}. The cardinality of the $\mathcal{VT}_{0}(n)$ code is at least $\frac{2^n}{n+1}$;
for the  exact cardinality of the $\mathcal{VT}_0(n)$ code, see \cite[Eq.~(10)]{Sloane01onsingle-deletion-correcting}.

For all $n$, the union of all VT codes forms a partition of the space $\FTwo^n$, that is $\bigcup_{a=0}^n \mathcal{VT}_a(n) = \FTwo^n$.

\section{Johnson-like Upper Bound}\label{sec:johnson}
In this section, we derive a Johnson-like upper bound on the list size which holds for \emph{any} $q$-ary code, minimum Levenshtein distance $d$, and any received word of length $N \geq 0$. Thus, from a combinatorial point of view, polynomial-time list decoding is possible at least up to the derived radius.

The proof follows similar ideas as the proof of Bassalygo \cite{Bassalygo1965New} and Roth~\cite[Prop.~4.11]{Roth_IntroductiontoCodingTheory_2006} 
for the Johnson bound in Hamming metric, but the adaptations to the Levenshtein metric are non-trivial since we have to deal with sequences of different lengths and spheres in the Levenshtein metric behave completely different than spheres in the Hamming metric. E.g., the size of a deletion sphere depends on its center, not only its radius.

\begin{theorem}[Main Theorem: Johnson-like Upper Bound]\label{thm:johnson_upper}
Let $ q \geq 2$ and let an integer $\tau <d \leq n+N$ be given. 
Let $\r \in \Fq^N$ be a given word of length $N$, where $N \in [n-\tau, n+\tau]$. 
Denote 
\begin{equation*}
\eta \triangleq \frac{q}{q+1} (n+N).
\end{equation*}
Then, for any $\codeID{n,M,d}$ code $\mycode{C}$
and for any integer $\tau$ such that 
\begin{equation}\label{eq:tau_C_johnson_qary}
\tau <   \eta - \sqrt{\eta(\eta-d)} \triangleq\tauID(q),
\end{equation}
the maximum list size $\ell= \max_{\vec{r} \in \Fq^{N}}\big\{\big|\mycode{C} \cap \BallID{\tau}{\vec{r}}\big|\big\}$ is bounded by
\begin{align}
\ell 
&\leq   \frac{d \eta}{\tau^2-(2\tau-d)\eta}\nonumber\\ 
& = \frac{d(n+N)}{\frac{q+1}{q} \tau^2 - (2\tau-d)(n+N)}.\label{eq:qary_johnsonbound}
\end{align} 
\end{theorem}

\begin{IEEEproof}
We denote the list of codewords as follows: 
$$\myset{L} \triangleq \mycode{C} \cap \BallID{\tau}{\r} = \{\c_1,\c_2,\dots,\c_{\ell}\} \subseteq \mycode{C} \subseteq \Fq^{n},$$
with $|\myset{L}|=\ell$.
Therefore, for each $i=1,\dots,\ell$, the number of insertions $\epsilon_i$ and the number of deletions $\delta_i$ of elements of $\vec{c}_i$ to obtain~$\vec{r}$ (where $\delta_i + \epsilon_i$ is minimized), satisfies $\epsilon_i + \delta_i= \dID(\vec{r},\vec{c}_i) \leq \tau$. Equivalently, $\dID(\vec{r},\vec{c}_i)$ is obtained by $\delta_i$ deletions of elements of $\vec{c}_i$ and $\epsilon_i$ deletions of elements of $\vec{r}$ such that both result in the same vector and $\delta_i + \epsilon_i$ is minimized.

We therefore denote sets of positions of deletions from~$\vec{c}_{i}$ and $\vec{r}$, respectively, of smallest size to obtain the same vector by $\myset{D}^{(i)}$ and $\myset{E}^{(i)}$, where $|\myset{D}^{(i)}| = \delta_i$ and $|\myset{E}^{(i)}| = \epsilon_i$. 
Note that these sets do not have to be unique, but the choice is fixed for the remainder of the proof.
We associate each $\vec{c}_{i}$ with the two corresponding sets~$\myset{D}^{(i)}$ and~$\myset{E}^{(i)}$. 
For these sets, the following holds: 
$$\myset{D}^{(i)} \subseteq [1,n] \text{ and }\myset{E}^{(i)} \subseteq [1,N].$$
The Levenshtein distance is therefore: $$\dID(\vec{r}, \vec{c}_i) =  |\myset{D}^{(i)}|+|\myset{E}^{(i)}| =  \delta_i + \epsilon_i \leq  \tau, \quad \forall i  \in [1,\ell].$$

Further, we denote by $\myset{F}^{(i_1,i_2)} \subseteq [1,n]$ and $\myset{F}^{(i_2,i_1)} \subseteq [1,n]$ the sets of positions which have to be deleted to obtain the same vector from $\vec{c}_{i_1}$ and from $\vec{c}_{i_2}$ such that the sum of their cardinalities $|\myset{F}^{(i_1,i_2)}| + | \myset{F}^{(i_2,i_1)}|$ is minimal. 
Therefore,
$$\dID(\vec{c}_{i_1}, \vec{c}_{i_2}) = |\myset{F}^{(i_1,i_2)}|+|\myset{F}^{(i_2,i_1)}| \geq d, \forall i_1, i_2  \in [1,\ell], i_1 \neq i_2.$$

Let $x_{a,j}$ denote the number of times in all $\ell$ codewords in~$\myset{L}$ that a fixed position $j \in [1,n]$ is contained in $\myset{D}^{(i)}$ and the deleted value equals $a \in \Fq$. 
Similarly, let $y_{a,j}$ denote the number of times in all $\ell$ codewords in $\myset{L}$ that $j \in [1,N]$ is contained in $\myset{E}^{(i)}$ and the deleted value equals $a \in \Fq$.

Thus,
\begin{equation}\label{eq:johnson_upperelltau}
\sum\limits_{a \in \Fq}^{} \Big(\sum\limits_{j\in[1,n]}  x_{a,j} +\sum\limits_{j\in[1,N]}y_{a,j} \Big)
\leq \ell \tau.
\end{equation}


Denote by $x_{\otimes,j}$ and $y_{\otimes,j}$ the number of times that a fixed position $j \notin \myset{D}^{(i)}$, respectively $j \notin \myset{E}^{(i)}$, $\forall i\in [1,\ell]$.
Hence,
\begin{equation*}
\sum\limits_{j \in [1,n]}^{} x_{\otimes,j} + \sum\limits_{j \in [1,N]}^{}y_{\otimes,j} \geq \ell(N+n-\tau).
\end{equation*}

In the following, consider all $\ell(\ell-1)$ tuples of two distinct codewords in $\mathcal{L}$, i.e., all tuples $(\vec{c}_{i_1},\vec{c}_{i_2})$ where $i_1 \neq i_2$, and their associated sets $\myset{D}^{(i_1)}, \myset{E}^{(i_1)}, \myset{D}^{(i_2)}, \myset{E}^{(i_2)}$.

We count the overall number of the following events $\forall i_1,i_2\in [1,\ell]$, $i_1 \neq i_2$, $\forall k \in [1,n]$ and $\forall j \in [1,N]$:
\begin{itemize}
	\item $k \in \myset{D}^{(i_1)} \cap \myset{D}^{(i_2)}$ and for the deleted values $a_{i_1} \neq a_{i_2}$
	\item or $k \notin \myset{D}^{(i_1)}$ and $k \in \myset{D}^{(i_2)}$
	\item or $k \in \myset{D}^{(i_1)}$ and $k \notin \myset{D}^{(i_2)}$
	\item or $j \notin \myset{E}^{(i_1)}$ and $j \in \myset{E}^{(i_2)}$
	\item or $j \in \myset{E}^{(i_1)}$ and $j \notin \myset{E}^{(i_2)}$.
\end{itemize} 
Notice that $$\{j \in \myset{E}^{(i_1)} \cap \myset{E}^{(i_2)} \text{ and for the deleted values } a_{i_1} \neq a_{i_2}\} = \emptyset,$$ since deleting the symbol at the same position of $\vec{r}$ always has the same value, otherwise this case would have to be considered as well.

The total number $\lambda$ of the previously mentioned five events can be bounded from below as follows:
\begin{align}
&  \lambda\triangleq \!\!\!\!\!\!\sum\limits_{i_1 \in [1,\ell]}\!\sum\limits_{\substack{i_2\in [1,\ell]\\ i_2 \neq i_1}}\!\!\!\! |\myset{D}^{(i_1)}\!\! \cup \myset{D}^{(i_2)} \!\setminus\! \{k:\! k\! \in \!\myset{D}^{(i_1)}\! \cap \myset{D}^{(i_2)}\!\! \wedge a_{i_1}\! \! =\! a_{i_2}\}|\nonumber\\
&\hspace{11ex} + |(\myset{E}^{(i_1)} \cup \myset{E}^{(i_2)})\setminus(\myset{E}^{(i_1)}\! \cap \myset{E}^{(i_2)})|\nonumber\\[2ex]
& \qquad\geq \sum\limits_{i_1\in [1,\ell]}\sum\limits_{\substack{i_2\in [1,\ell]\\ i_2 \neq i_1}} |\myset{F}^{(i_1,i_2)}|+|\myset{F}^{(i_2,i_1)}|
\geq \ell (\ell-1)d.\label{eq:count-events}
\end{align}

On the other hand, we can explicitly count the five events as follows: given a position $k \in [1,n]$ and a fixed element $a \in \Fq$, the list size equals $\ell = x_{a,k} + \hat{x}_{a,k} + x_{\otimes,k}$ where $\hat{x}_{a,k}$ denotes the number of values $i$, $i \in [1,\ell]$, for which $k \in \mathcal{D}^{(i)}$ but where the deleted $k$-th symbol of $\vec{c}^{(i)}$ does not equal $a$. Then, the number of values $i_1$ and $i_2$, $i_1,i_2 \in [1,\ell]$ and $i_1 \neq i_2$, for which the first event holds (for a fixed $k$) is $\sum_{a \in \Fq} x_{a,k} \cdot \hat{x}_{a,k}$. Similarly, the number of pairs $(i_1,i_2)$ for which the second or third event holds are $\sum_{a \in \Fq} x_{a,k} \cdot {x}_{\otimes,k} + x_{\otimes,k}(\ell-x_{\otimes,k})$. Thus, summing over the number of pairs $(i_1,i_2)$ for which one of the first three events occur for a fixed $k$, gives $\sum_{a \in \Fq \cup \otimes} x_{a,k} (\ell - x_{a,k})$.

A similar argument shows that the total number of pairs $(i_1,i_2)$ of the last two events for a fixed $j$ equals $\sum_{a \in \Fq \cup \otimes} y_{a,j} (\ell - y_{a,j})$. It therefore follows from summing over all $j$ and $k$ that (pay attention that the $\otimes$ is included in the summation):
\begin{equation}
\lambda = \sum\limits_{a \in \Fq \cup \otimes}\Big( \sum\limits_{k\in [1,n]} x_{a,k}(\ell-x_{a,k}) +\sum\limits_{j\in [1,N]}y_{a,j} (\ell-y_{a,j})\Big).\label{eq:def_lambda}
\end{equation}
%
%
%
%
%
Since $\sum_{a \in \Fq \cup \otimes} x_{a,k} = \ell$ for all $k\in [1,n]$ and $\sum_{a \in \Fq \cup \otimes} y_{a,j} = \ell$ for all $j\in [1,N]$, we can simplify~\eqref{eq:def_lambda} as follows: 
\begin{equation}
\lambda = \ell^2(n+N) - \sum\limits_{a \in \Fq \cup \otimes} \Bigg(\sum\limits_{k\in [1,n]} x_{a,k}^2 + \sum\limits_{j\in [1,N]}y_{a,j}^2\Bigg).\label{eq:johnson_equaldiff}
\end{equation}
Combining \eqref{eq:count-events} and \eqref{eq:johnson_equaldiff} yields:
\begin{equation*}
\ell^2(n + N)-\ell(\ell-1)d 
\geq \!\!\sum\limits_{a \in \Fq \cup \otimes} \Bigg(\sum\limits_{k\in [1,n]} x_{a,k}^2 + \sum\limits_{j\in [1,N]}y_{a,j}^2\Bigg).
\end{equation*}
Further, $x_{\otimes,k} =\ell -\sum_{a \in \Fq} x_{a,k}$ and similarly, $y_{\otimes,j} =\ell -\sum_{a \in \Fq} y_{a,j}$, and therefore, 
$$x_{\otimes,k}^2 =\Big(\ell -\sum_{a \in \Fq}x_{a,k}\Big)^2 \quad \text{and} \quad y_{\otimes,j}^2 =\Big(\ell -\sum_{a \in \Fq}y_{a,j}\Big)^2.$$
Hence,
\begin{align}
&\ell^2(n+N)-\ell(\ell-1)d \nonumber\\
&\geq  \sum\limits_{a \in \Fq}\! \left(\sum\limits_{k\in [1,n]} x_{a,k}^2 +\! \!\!\sum\limits_{j\in [1,N]}y_{a,j}^2\!\right)
\!+\!\!\! \sum\limits_{k\in [1,n]} x_{\otimes,k}^2\! +\!\!\! \sum\limits_{j\in [1,N]}y_{\otimes,j}^2\nonumber\\[1ex]
& = \!\ell^2(n\!+\!N)
+\!\!\!\! \sum\limits_{k\in [1,n]}\!\! \Bigg(\!\sum_{a \in \Fq} x_{a,k}^2\! - 2 \ell \sum_{a \in \Fq}\!x_{a,k} \!+ \!\Big(\!\sum_{a \in \Fq}x_{a,k} \Big)^{\!2}\!\Bigg)
\nonumber\\
&\hspace{3ex} + \sum\limits_{j\in [1,N]} \Bigg(\sum_{a \in \Fq} y_{a,j}^2 - 2 \ell \sum_{a \in \Fq}y_{a,j} + \Big(\sum_{a \in \Fq}y_{a,j} \Big)^2\Bigg).\label{eq:johnson_comb}
\end{align}

The Cauchy--Schwarz inequality states that 
$$\sum_{a \in \Fq}x_{a,k}^2 \geq \frac{1}{q}\cdot \Big(\sum_{a \in \Fq}x_{a,k}\Big)^2.$$ 
Further, for ease of notation, we denote:
\begin{equation*}
X_k \triangleq \sum_{a \in \Fq} x_{a,k}, \quad \text{and} \quad Y_j \triangleq \sum_{a \in \Fq} y_{a,j}.
\end{equation*}

Plugging this into~\eqref{eq:johnson_comb} gives:
\begin{align}
\ell&(\ell-1)d \nonumber\\
 &\leq \! \sum\limits_{k\in [1,n]}\!\!\left(\! 2\ell X_k- \frac{q+1}{q}X_k^2  \right)\!+\!\! \sum\limits_{j\in [1,N]}\!\left(2\ell Y_j - \frac{q+1}{q} Y_j^2\! \right)\!.\label{eq:johnson_proof_maximize}
\end{align}
We will now bound the RHS of~\eqref{eq:johnson_proof_maximize} from above. 
For that purpose, denote the following function $f(z)$ (over the reals):
\begin{equation*}
f(z) = 2\ell z - \frac{q+1}{q} z^2.
\end{equation*}
This function $f(z)$ is monotonically increasing for $z \leq \frac{q}{q+1} \ell$, achieves its global maximum for $z_{max}\triangleq z  = \frac{q}{q+1} \ell$ where $f(z_{max}) = \ell^2 \frac{q}{q+1}$.
For bounding \eqref{eq:johnson_proof_maximize}, we have to take into account the restriction from \eqref{eq:johnson_upperelltau}, i.e.,
\begin{equation}\label{eq:fct-restriction}
\sum\limits_{k\in[1,n]}  X_{k} +\sum\limits_{j\in[1,N]}Y_{j}
\leq \ell \tau.
\end{equation}
Since $f(X_k)$ and $f(Y_j)$ are monotonically increasing, also the following sum (which is equal to the RHS of~\eqref{eq:johnson_proof_maximize})
$$\sum\limits_{k\in[1,n]}  f(X_{k}) +\sum\limits_{j\in[1,N]} f(Y_{j})$$
is monotonically increasing and has its maximum value when $X_k = Y_j = \frac{q}{q+1} \ell$ which is $\frac{q}{q+1} \ell^2 (n+N)$.
Since $\ell \tau \leq \frac{q}{q+1} \ell^2 (n+N)$, the RHS of \eqref{eq:johnson_proof_maximize} is maximized under the restriction of \eqref{eq:fct-restriction} for 
$$X_k = Y_j = \frac{\ell \tau}{n+N}, \qquad \forall i \in [1,n], k \in [1,N].$$
Therefore, we can bound~\eqref{eq:johnson_proof_maximize} from above as follows:
\begin{equation}\label{eq:johnson_laststep}
\ell(\ell-1)d \leq 2 \ell^2 \tau - \frac{q+1}{q} \frac{\ell^2\tau^2}{n+N}.
\end{equation}
Solving \eqref{eq:johnson_laststep} for $\ell$ leads to
the statement of Theorem~\ref{thm:johnson_upper}. The restriction on $\tau$ holds since the denominator of \eqref{eq:qary_johnsonbound} has to be positive.
\end{IEEEproof}

The following corollary shows an alphabet-free Johnson-like upper bound which holds since the bound from~\eqref{eq:qary_johnsonbound} can be bounded from above by~\eqref{eq:noq_johnsonbound}.
\begin{corollary}[Alphabet-free Johnson-like Bound]\label{cor:johnson_upper}
	For any $\codeID{n,M,d}$ code $\mycode{C}$, any received word of length $N$,
	and any integer $\tau$ such that 
	\begin{equation}\label{eq:def_tauCC}
	\tau < 
	n+N - \sqrt{(n+N)(n+N-d)} \triangleq \tauID,
	\end{equation}
	the list size $\ell$ is bounded from above by
	\begin{equation}
	\ell
	\leq 
	\frac{(n+N)d}{\tau^2-(2\tau-d)(n+N)}\triangleq \ellID.\label{eq:noq_johnsonbound}
	\end{equation} 
\end{corollary}

We have therefore derived an upper bound on the list size, depending on the length of the code $n$, the minimum Levenshtein distance of the code $d$, and the length of the received word $N$. This bound holds for any number of insertions $\epsilon$ and number of deletions $\delta$ where $\epsilon + \delta \leq \tau$, when $\tau$ satisfies~\eqref{eq:tau_C_johnson_qary}, respectively \eqref{eq:def_tauCC}.
Thus, polynomial-time list decoding is feasible up to this radius. 

Corollary~\ref{cor:johnson_upper} is an alphabet-free version of the Johnson-like bound from Theorem~\ref{thm:johnson_upper} which is clearly weaker.
Similar to the Johnson bound in Hamming metric, taking into account the field size $q$ makes the largest difference for the smallest field size $q$, i.e., for $q=2$.

\section{Discussion of the Johnson-like Bound}\label{sec:discussion}
\subsection{Illustration}
In this section, we plot our Johnson-like bounds in order to illustrate the gain in the decoding radius compared to unique decoding.

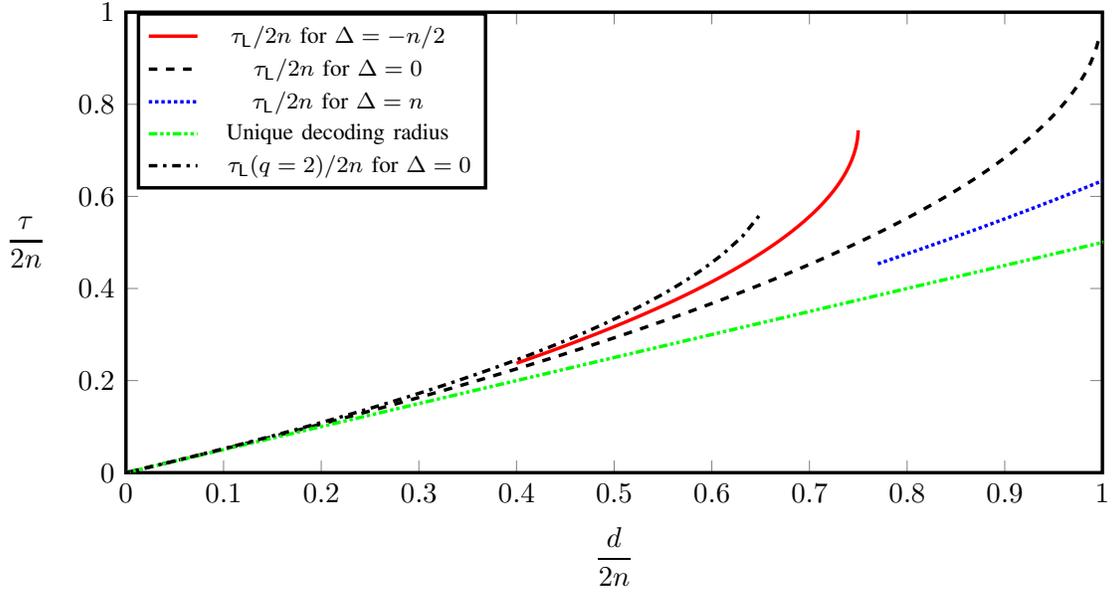
\begin{figure*}[htb]
	\centering
	\tikzsetnextfilename{decoding_radii}
	\begin{tikzpicture} [scale = 1.08]
	\begin{axis}
	[
	legend style={font=\footnotesize, at={(0.37,1)}},
	xlabel = {$\dfrac{d}{2n}$},
	ylabel = {$\dfrac{\tau}{2n}$},
	ylabel style = {rotate=270},
	xmin = 0,
	xmax = 1,
	ymin = 0,
	ymax = 1,
	line width = 1.2,
	width = 0.75\textwidth,
	height = 0.4\textwidth
	]
	\addplot[red, solid, samples=2000,domain = 0.4:1]{0.75-sqrt(9/16-0.75*x)}; 
	\addplot[black,dashed, samples=2000]{1-sqrt(1-x)}; 
	\addplot[blue, densely dotted,samples=500,domain = 0.77:1]{1.5-sqrt(2.25-1.5*x)}; 
	\addplot[green,densely dash dot dot]{0.5*x}; 
	\addplot[black, dash dot,samples=500]{2/3-sqrt(4/9-2*x/3)}; 
	\legend{$\tauID/2n$ for $\Delta = -n/2$, $\tauID/2n$ for $\Delta = 0$, $\tauID/2n$ for $\Delta = n$, Unique decoding radius, $\quad\tauID(q=2)/2n$ for $\Delta=0$};
	\end{axis}
	\end{tikzpicture}
	\caption{Normalized alphabet-free Johnson-like radius $\frac{\tauID}{2n}$ from~\eqref{eq:def_tauCC} (first three curves), as a function of the normalized minimum Levenshtein distance $\frac{d}{2n}$. The unique decoding radius (i.e., $\frac{\tau}{2n} = \frac{d}{4n}$) corresponds to the green dash-dotted line.	
	The last curve (black dash dotted) illustrates the alphabet-dependent normalized Johnson-like radius $\frac{\tauID(q)}{2n}$ from~\eqref{eq:tau_C_johnson_qary} for $q=2$ and $\Delta = 0$.
		\vspace{1ex}
		\label{fig:bounds}}
	\hrulefill
\end{figure*}

First, we illustrate the alphabet-free Johnson-like upper bound (Corollary~\ref{cor:johnson_upper}).
Since $\Delta \triangleq N-n$ depends on $\tau$, we normalize the radius by $2n$ (and not by $n+N$) and illustrate $\frac{\tauID}{2n}$ of~\eqref{eq:def_tauCC} as a function of $\frac{d}{2n}$ for different values of $\Delta$ in Fig.~\ref{fig:bounds}. 
This normalization of~\eqref{eq:def_tauCC} gives:
\begin{equation}\label{eq:Johnson-like-normalized}
\frac{\tauID}{2n}  = 1+\frac{\Delta}{2n} - \sqrt{ 1+\frac{\Delta}{n} +\frac{\Delta^2}{4n^2} - \left(1+\frac{\Delta}{2n}\right) \frac{d}{2n}}
\end{equation} 
Fig.~\ref{fig:bounds} illustrates~\eqref{eq:Johnson-like-normalized} for different values of $\Delta$. 

Notice that since $N \in [n-\tau, n+\tau]$, we know that $\Delta \in [-\tau,\tau]$.
Thus, for e.g., $\Delta = -\frac{n}{2}$, we have $\frac{\tau}{2n} \geq \frac{1}{4}$. On the other hand, the argument in the square root is only defined for $\frac{\tau}{2n} \leq \frac{3}{4}$ and therefore the curve is only plotted for this region.

In general, the plot shows a notable improvement compared to the unique decoding radius~\eqref{eq:unique-dec} (green {dash-dotted} line). 

Interestingly, for $\Delta = 0$, the result equals the alphabet-free Hamming-metric Johnson radius $n-\sqrt{n(n-d)}$ when normalized by $n$.

Second, we take the field size into account which gives a further improvement, especially for small $q$. 
The alphabet-dependent bound from~\eqref{eq:tau_C_johnson_qary} when normalized by $2n$ gives:
\begin{align*}
\frac{\tauID(q)}{2n}  =& \frac{q}{q+1}\left(1+\frac{\Delta}{2n}\right)\\ &-\! \sqrt{ \frac{q}{q+1}\!\left(1+\frac{\Delta}{2n}\right)\cdot\left(\frac{q}{q+1}\left(1+\frac{\Delta}{2n}\right)\!-\!\frac{d}{2n}\right) }.
\end{align*} 
For $q=2$, this gives:
\begin{equation}\label{eq:Johnson-like-normalized-q2}
\frac{\tauID(q)}{2n}  = \frac{2}{3} + \frac{\Delta}{3n} - \sqrt{\frac{4}{9} + \frac{4\Delta}{9n}+\frac{\Delta^2}{9n^2}-\left(\frac{2}{3}+\frac{\Delta}{3n}\right)\frac{d}{2n}}.
\end{equation} 
In Fig.~\ref{fig:bounds}, the last black dash-dotted curve illustrates $\frac{\tauID(q)}{2n}$ for $\Delta = 0$. Compared to the second curve, we notice that taking the field size into account demonstrates a further improvement.

\subsection{VT Codes}
The following example shows that polynomial-time list decoding of two insertions/deletions is possible with VT codes.
\begin{example}[Johnson-like Bound for VT Codes]\label{ex:VT-Codes}
	First note that VT codes have Levenshtein distance $d = 4$
	since all codewords have equal length $n$ (thus $d$ has to be even) and since it is proven that they can correct any single deletion or insertion (and therefore $d \geq 3$). 
	
	For fixed $d$, the larger $n+N$ gets, the smaller the RHS of~\eqref{eq:tau_C_johnson_qary} and \eqref{eq:def_tauCC} get. Thus, for fixed $n$ and $d$ and $N \in \intervallincl{n-\tau}{n+\tau}$, the minimum of the RHS is achieved for $N = n + \tau$. Thus, if we prove that $\tau$ insertions can be corrected, any combination of $\tau$ insertions and deletions can be corrected as well.
	
	We want to understand if list decoding of VT codes for $\tau = 2$ is possible for any $N \in \intervallincl{n-2}{n+2}$ and therefore check \eqref{eq:def_tauCC} for $d = 4$ and $N = n+2$.
	From \eqref{eq:def_tauCC}, we therefore need that:
	\begin{align*}
	2n+2 - \sqrt{(2n+2)(2n-2)} &>2.\\
	\Longleftrightarrow 2n = \sqrt{4n^2} > \sqrt{(2n-2)(2n+2)}& = \sqrt{4n^2-4},
	\end{align*}
	which is true for any $n>0$.
	
	Thus, list decoding of two insertions/deletions is feasible in polynomial time.
	This potentially doubles the error-correcting capability of VT codes (at the cost of having a list of codewords). In Section~\ref{sec:list-dec-vt}, a simple list decoding algorithm for this case is shown and also how to generalize it to any constant number of insertions/deletions.
\end{example}
\subsection{Only Deletions or Only Insertions}
The bound from Theorem~\ref{thm:johnson_upper} simplifies to the following corollary when only deletions occur.

\begin{corollary}[Johnson-like Upper Bound for Deletions]\label{cor:johnson_upper_del}
Let $ q \geq 2$  and
let $\r \in \Fq^N$ be a given word of length $N = n- \delta \geq 0$.
Then, for any $\codeID{n,M,d}$ code $\mycode{C}$
and for any integer $\delta$ such that 
\begin{equation*}
\frac{q+1}{q} \delta^2 - (2\delta-d)(2n-\delta) >0,
\end{equation*}
the list size $\ell$ is bounded from above by
\begin{align*}
\ell&= \max_{\vec{r} \in \Fq^{N}}\Big\{\big|\mycode{C} \cap \BallD{\delta}{\vec{r}}\big|\Big\} \nonumber\\
& \leq \frac{d(2n-\delta)}{\frac{q+1}{q} \delta^2 - (2\delta-d)(2n-\delta)},
\end{align*} 
where $\BallD{\delta}{\vec{r}}$ denotes a deletion ball around $\vec{r}$ of deletion radius~$\delta$, i.e., the number of deletions $\delta$ between all words in the ball and $\vec{r}$ is at most $\delta$.
\end{corollary}

Similarly, we obtain  the following Johnson-like upper bound for list decoding only insertions.
\begin{corollary}[Johnson-like Upper Bound for Insertions]\label{cor:johnson_upper_ins}
Let $ q \geq 2$ and 
let $\r \in \Fq^N$ be a given word of length $N = n+ \epsilon \geq n$.
Then, for any $\codeID{n,M,d}$ code $\mycode{C}$
and for any integer $\epsilon$ such that 
\begin{equation*}
\frac{q+1}{q} \epsilon^2 - (2\epsilon-d)(2n+\epsilon) >0,
\end{equation*}
the list size $\ell$ is bounded from above by
\begin{align*}
\ell&= \max_{\vec{r} \in \Fq^{N}}\Big\{\big|\mycode{C} \cap \BallIn{\epsilon}{\vec{r}}\big|\Big\} \nonumber\\
& \leq \frac{d(2n+\epsilon)}{\frac{q+1}{q} \epsilon^2 - (2\epsilon
	-d)(2n+\epsilon)},
\end{align*} 
where $\BallI{\epsilon}{\vec{r}}$ denotes a ball around $\vec{r}$ of insertion radius~$\epsilon$, i.e., the number of insertions $\epsilon$ between all words in the ball and $\vec{r}$ is at most $\epsilon$.
\end{corollary}

Interestingly, the Johnson-like bound for deletions is larger than the one for insertions; i.e., it seems that more deletions can be list decoded than insertions.

\section{A Lower Bound for Deletion-List-Decoding of VT Codes}\label{sec:lower-bound}
In this section, we derive a lower bound on the maximum list size when list decoding \emph{deletions only} with VT codes.
In general, a lower bound on the maximum list size will give a minimum complexity estimate of any list decoder since stating the list of codewords is a necessary task of any list decoder.

\begin{theorem}[Lower Bound of VT Codes for Deletions]\label{thm:lower-deletions}
	For the binary $\mathcal{VT}_0(n)$ code of cardinality $|\mathcal{VT}_0(n)| \geq \frac{2^n}{n+1}$, the maximum list size of list decoding up to $\delta$ deletions is bounded from below as follows:
	\begin{equation*}
	\ell \geq  \frac{\binom{n}{\delta}}{n+1} \geq \left(\frac{n}{\delta}\right)^{\delta}\cdot\frac{1}{n+1} \gtrsim \frac{n^{\delta-1}}{\delta^{\delta}}.
	\end{equation*}
\end{theorem}
\begin{IEEEproof}
	The length of the received word $\vec{r}$ is $N = n-\delta$.
	Due to the pigeonhole principle, there is a received word $\vec{r}$ of length $N$ such that the number of codewords of length~$n$ around $\vec{r}$ in insertion radius \emph{at most} $\delta$ is at least the average number of codewords in an insertion ball of radius $\delta$ (whose size does not depend on $\vec{r}$, but on $N$):
	\begin{align*}
	\ell \geq |\mycode{C}| \cdot \frac{|\BallIn{\delta}{N}|}{2^n}.
	\end{align*}
	It is known that for a binary vector of length $N$, the number of supersequences of length $N+s$ is exactly $\sum_{i=0}^{s} \binom{N+s}{i}$.
	Thus, the number of supersequences of a vector of length $N$ of length \emph{at most} $N+\delta$ is $|\BallIn{\delta}{N}| = \sum_{s=0}^{\delta}\sum_{i=0}^{s} \binom{N+s}{i}$ and with the size of the VT code, we obtain:
	\begin{align*}
	\ell &\geq \frac{|\BallIn{\delta}{N}|}{n+1}
	 = \frac{\sum_{s=0}^{\delta}\sum_{i=0}^{s} \binom{N+s}{i}}{n+1} 
	\end{align*}	
	and the statement follows.
\end{IEEEproof}

The maximum list size therefore grows exponentially with the number of deletions and at least polynomially with a degree-$\delta$-function in the length $n$ of the VT code.
Notice that this result does not contradict in any sense Example~\ref{ex:VT-Codes} or Algorithm~\ref{algo:vt-list-decoder}, since this shows in general a lower bound on list decoding $\delta$ deletions, while Example~\ref{ex:VT-Codes} and Algorithm~\ref{algo:vt-list-decoder} show that list decoding of a \emph{constant} number deletions with VT codes always results in a polynomial-sized list of codewords. 

Deriving a similar lower bound for list decoding of \emph{insertions} does not work in this way since the actual size as well as lower bounds on the size of the deletion ball $|\BallD{\delta}{\vec{r}}|$ depend on the received word rather than only its length (or actually, the number of runs of the received word), see~\cite[Equation~(1)]{Levenshtein-binarycodesCorrectingDeletions} and therefore the pigeonhole argument of the proof of Theorem~\ref{thm:lower-deletions} does not work.
The same problem occurs when deriving a lower bound for insertions \emph{and} deletions. Additionally, we are not aware of an expression or an explicit lower bound for $|\BallID{\tau}{\vec{r}}|$.

\section{List Decoding Algorithm for VT Codes}\label{sec:list-dec-vt}
As shown in Example~\ref{ex:VT-Codes}, polynomial-time list decoding of VT codes for $\tau=2$ is possible. In this section, we present a simple $(2,n)_\mathsf{L}$-list decoder for VT codes. We restrict ourselves to $a=0$, but this principle works for any other VT code as well. Usually, the list size will be much smaller than~$n$. Despite its simple concept, this decoder is the first explicit polynomial-time list decoding algorithm in the Levenshtein metric. It is also shown how to generalize the algorithm to any constant number of insertions/deletions.

We denote the following \emph{checksum} $S(\vec{r})$ of the received word $\vec{r}$ (which is zero if $\vec{r} \in \mathcal{VT}_0(n)$ and $N=n$):
\begin{equation}\label{eq:checksum}
S(\vec{r}) \triangleq -\sum_{i=1}^{N} i \cdot r_i \bmod (n+1).
\end{equation}
Notice that the minus for the checksum is introduced such that~\eqref{eq:checksum} rather denotes the checksum \emph{deficiency} as in \cite{Sloane01onsingle-deletion-correcting}.

The length of the received word $\vec{r} = (r_1,r_2,\dots,r_N)$ is denoted by $N$ and provides information on how many insertions/deletions occurred. We thereby assume that the total number of insertions and deletions is at most $\tau = 2$. Thus, for $N=n-2$, we will correct two deletions; for $N=n-1$ we correct a single deletion (with list size one); for $N=n$ we will correct one insertion and one deletion (if the received word is already a codeword, the list size will be one as explained later); for $N=n+1$, one insertion; and for $N=n+2$, two insertions.

For our list decoder, we make use of the well-known single insertion/deletion decoder for VT codes, cf. \cite{Sloane01onsingle-deletion-correcting}, which is summarized in Algorithm~\ref{algo:vt-decoder} for the case when $N = n-1$, i.e., a single deletion happened.
Here, from a VT codeword $(c_1,c_2,\dots,c_n)$ the symbol $c_p$ was deleted. 
Denote by $L_0$/$L_1$ the number of zeros/ones left of $c_p$ and by $R_0$/$R_1$ the number zeros/ones right of $c_p$. Note that $\wtH{\vec{r}} = R_1 + L_1$. 
Of course, the decoder knows neither $p$ nor $c_p$ nor $L_0, L_1, R_0, R_1$.

\printalgoIEEE{
	\caption{
		\textsc{VT-Del-Dec($\vec{r}$)}}
	\label{algo:vt-decoder}
	\DontPrintSemicolon
	\SetAlgoVlined
	\SetSideCommentRight
	\LinesNumbered
	\SetKwInput{KwIn}{\underline{Input}}
	\SetKwInput{KwOut}{\underline{Output}}
	\SetKwInput{KwIni}{\underline{Initialize}}
	\KwIn{Received sequence: $\vec{r}=(r_1, \dots, r_{n-1}) \in \mathbb{F}_{2}^{n-1}$}
	\BlankLine
	Weight calculation: $w = \wtH{\vec{r}}$\\
	Checksum calculation: $S(\vec{r})$\\
	\If{$S(\vec{r})\leq w$}{
		Find $p$ such that $\wtH{r_p,\dots,r_{n-2},r_{n-1}} = S(\vec{r})$\\
		\KwOut{Reconstructed sequence of length $n$: $(r_1,\dots,r_{p-1},\textcolor{blue}{0}, r_p,\dots,r_{n-1})$}
		}
	\Else{$L_0 = S(\vec{r})-w-1$\\
		Find $p$ such that $(r_1,r_2,\dots,r_{p-1})$ contains $L_0$ zeros\\
		\KwOut{Reconstructed sequence of length $n$: $(r_1,\dots,r_{p-1},\textcolor{blue}{1}, r_p,\dots,r_{n-1})$}	
		}
}
The complexity of Algorithm~\ref{algo:vt-decoder} is in $\mathcal{O}(n)$.
Single \emph{insertion}-correction works in a similar way and is denoted by \textsc{VT-In-Dec} in the following. 

Our list decoding algorithm uses the VT single insertion/deletion decoder and is shown in Algorithm~\ref{algo:vt-list-decoder}. If $N=n-1$ or $n+1$, we directly call a unique VT decoder since two insertions or deletions in a vector of length $n$ cannot result in a vector of length $n-1$ or $n+1$. In the other cases, we first generate all subsequences or supersequences of length $n-1$ or $n+1$, respectively, and call a VT decoder for each of these sequences.

\printalgoIEEE{
	\caption{
		\textsc{VT-List-Dec($\vec{r}$)}}
	\label{algo:vt-list-decoder}
	\DontPrintSemicolon
	\SetAlgoVlined
	\SetSideCommentRight
	\LinesNumbered
	\SetKwInput{KwIn}{\underline{Input}}
	\SetKwInput{KwOut}{\underline{Output}}
	\SetKwInput{KwIni}{\underline{Initialize}}
	\KwIn{Received sequence: $\vec{r}=(r_1, r_2, \dots, r_{N}) \in \mathbb{F}_{2}^{N}$}
	\BlankLine
	$\mathcal{L} = \emptyset$\\
	\If{$N=n-2$}{
		\For{all supersequences $\vec{r}^\prime$ of $\vec{r}$ of length $n-1$}
		{$\vec{c} = $\textsc{VT-Del-Dec}($\vec{r}^\prime$)\\
		\If{$S(\vec{c}) = 0$ and $\dID(\vec{r},\vec{c})\leq 2$}{Add $\vec{c}$ to $\mathcal{L}$}	}
		}
	\If{$N=n-1$}{
		$\vec{c} = $\textsc{VT-Del-Dec}($\vec{r}$)\\
		$\mathcal{L} = \{\vec{c}\}$}
	\If{$N=n$}{
		
				\For{all subsequences $\vec{r}^\prime$ of $\vec{r}$ of length $n-1$}
						{$\vec{c} = $\textsc{VT-Del-Dec}($\vec{r}^\prime$)\\
							\If{$S(\vec{c}) = 0$ and $\dID(\vec{r},\vec{c})\leq 2$}{Add $\vec{c}$ to $\mathcal{L}$}	}
						
				\For{all supersequences $\vec{r}^\prime$ of $\vec{r}$ of length $n+1$}
				{$\vec{c} = $\textsc{VT-In-Dec}($\vec{r}^\prime$)\\
					\If{$S(\vec{c}) = 0$ and $\dID(\vec{r},\vec{c})\leq 2$}{Add $\vec{c}$ to $\mathcal{L}$}	}

		}
	\If{$N=n+1$}{$\vec{c} = $\textsc{VT-In-Dec}($\vec{r}$)\\
		$\mathcal{L} = \{\vec{c}\}$}
	\If{$N=n+2$}{
		\For{all subsequences $\vec{r}^\prime$ of $\vec{r}$ of length $n+1$}
		{$\vec{c} = $\textsc{VT-In-Dec}($\vec{r}^\prime$)\\
			\If{$S(\vec{c}) = 0$ and $\dID(\vec{r},\vec{c})\leq 2$}{Add $\vec{c}$ to $\mathcal{L}$}	}
		
		}
	\KwOut{List of possible codewords $\mathcal{L} \subset \Fq^n$}
}
Clearly, this strategy provides all codewords around $\vec{r}$ in distance at most $\tau=2$.
To understand the maximum list size, recall that for $N=n-2$, the number of supersequences of length $n-1$ is exactly $n$; for $N=n+2$, the number of subsequences of length $n+1$ is equal to the number of runs of $\vec{r}$ and therefore at most $n+2$.
Further, there cannot be two codewords in the output list which result in the given received word by deleting the symbol at the same position (and one additional insertion or deletion at different positions) since otherwise these two words have Levenshtein distance only two. Thus, the list size for $N=n+2$ and also for $N=n$ is at most $n$.

The previously mentioned reasoning explains also why the list size will be one if $\vec{r}$ has length $N=n$ and is a codeword (i.e., $S(\vec{r}) = 0$).

The complexity of Algorithm~\ref{algo:vt-list-decoder} is in $\mathcal{O}(n^2)$ since $\mathcal{O}(n)$ times a unique VT decoder is called.

This algorithm can be extended to any constant number of insertions and deletions $\tau$, by building simply all the corresponding sub- and supersequences. 
For example, when $N=n$, we know that the same number of insertions and deletions occurred; From $\vec{r}$ we then create 
\begin{itemize}
	\item all supersequences of length $n+1$ and use a single insertion decoder,
	\item all subsequences of length $n-1 $ and use a single deletion decoder,
	\item all supersequences of length $n+2$, from each of these sequences all subsequences of length $n+1$ and then use a single insertion decoder,
	\item ...
	\item all subsequences of length $n-\lfloor\tau/2\rfloor$, from each of these sequences all supersequences of length $n-\lfloor\tau/2\rfloor-1$, from each of these sequences all supersequences of length $n-\lfloor\tau/2\rfloor-2$ and so on until we have many sequences of length $n-1$ and use a single deletion decoder.
\end{itemize}
For analyzing the list size, let us look at the last step: here the number of subsequences of length $n-\lfloor\tau/2\rfloor$ is at most $n^{\lfloor\tau/2\rfloor}$, for each time when a new set of supersequences of these sequences is created, the number of investigated sequences is increased by a factor of $n$. Thus, a VT decoder has to be applied to at most $n^{\lfloor\tau/2\rfloor} \cdot n^{\lfloor\tau/2\rfloor-1} = n^{\tau-1}$ sequences. Since there are in total $\tau$ such steps, at most $\tau n^{\tau-1}$ sequences have to be investigated, i.e., for constant $\tau$, the list size is in $\mathcal{O}(n^{\tau-1})$. For $\tau=2$, we could previously show that the list size is at most $n$, and not $2n$, so it might be possible to prove that for the general case the list size is at most $n^{\tau-1}$ instead of $\tau n^{\tau-1}$, but this is not obvious.

 
In the case for a constant number $\delta$ of \emph{deletions only}, for any received word of length $N=n-\delta$, we have to calculate all supersequences of length $n-1$ and apply a VT decoder. In this case, the list size is at most $n^{\delta-1}$.

\section*{Acknowledgment}
The author would like to thank the reviewers whose very helpful comments helped to improve the paper significantly (and in particular to make the proof of the main result more understandable) and also the reviewers from ISIT 2017 (one of the reviewers made a comment which significantly simplified the list decoder of VT codes).

Further, the author wants to thank Andreas Lenz and Eitan Yaakobi for the helpful discussions.
\bibliographystyle{IEEEtranS}
\bibliography{antoniawachter}

\begin{IEEEbiographynophoto}{Antonia Wachter-Zeh}
	(S’10--M'14) is an Assistant Professor at the Technical University of Munich (TUM), Munich Germany.
	She received an equivalent of the B.S. degree in electrical engineering in 2007 from the University of Applied Science Ravensburg, Germany, and the M.S. degree
	in communications technology in 2009 from Ulm University, Germany.
	She obtained her Ph.D. degree in 2013 at the Institute of Communications Engineering, University of Ulm, Germany and at the 
	Institut de recherche mathématique de Rennes (IRMAR), Université de Rennes 1, Rennes, France. 
	From 2013 to 2016, she was a postdoctoral researcher at the Technion---Israel Institute of Technology, Haifa, Israel.
	Her research interests are coding and information theory and their application to storage, communications, and security.
\end{IEEEbiographynophoto}

\end{document}